# CHARACTERIZING USER BEHAVIOR AND INFORMATION PROPAGATION ON A SOCIAL MULTIMEDIA NETWORK

Francis T. O'Donovan[1], Connie Fournelle[1], Steve Gaffigan[1], Oliver Brdiczka[2], Jianqiang Shen[2], Juan Liu[2], and Kendra E. Moore[1]

[1]Boston Fusion Corp, 1 Van de Graaff Dr Ste 107, Burlington, MA 01803, USA
{francis.odonovan, connie.fournelle, stephen.gaffigan, kendra.moore}@bostonfusion.com
[2]Palo Alto Research Center, 3333 Coyote Hill Road, Palo Alto, CA 94304, USA
{oliver.brdiczka, jianqiang.shen, juan.liu}@parc.com

## ABSTRACT

An increasing portion of modern socializing takes place via online social networks. Members of these communities often play distinct roles that can be deduced from observations of users' online activities. One such activity is the sharing of multimedia, the popularity of which can vary dramatically. Here we discuss our initial analysis of anonymized, scraped data from consenting Facebook users, together with associated demographic and psychological profiles. We present five clusters of users with common observed online behaviors, where these users also show correlated profile characteristics. Finally, we identify some common properties of the most popular multimedia content.

## 1. INTRODUCTION

The explosion in the use of online social networks over the past decade (from 8% of all internet users in 2005 to 67% in 2012 [1]) suggests that these internet communities are, and will continue to be, an ever more dominant forum for social interactions. This increased use has been accompanied (and likely fueled) by the transition from text-only communications to a richer multimedia experience. Publicly accessible communities such as Facebook have developed alongside limited-access social platforms such as Yammer and Socialcast, which provide internal corporate networks for day-to-day business communication, and carry these social interaction methods into the workspace.

Throughout history, members of (offline) social communities have often taken on roles within those groups without these roles being formally recognized. For example, employees that act within companies as brokers of information perform one of the following roles [2]: (i) *coordinator* within a department; (ii) *consultant* external to the department; (iii) *gatekeeper* dealing with external requests; (iv) department *representative*; or (v) *liaison* between departments. In a similar way, users of online social networks take on such 'latent' roles – for example, are they mostly a creator or consumer of content? These roles are not specifically identified, and are separate from formal roles (such as company manager) that they may have on these networks. And when a social network user's latent role involves content creation, the number of consumers of this multimedia content can vary greatly. Most content is viewed by few other users, but some limited number of these multimedia broadcasts (online postings of content) go 'viral' and become extremely successful and popular, being viewed by hundreds of millions of users [3].

The field of Psychology provides tools to explain how a person behaves and what latent roles they adopt – whether offline or online. One of the most widely-adopted personality models is the Big Five Model [4] that considers five dominant personality traits (namely Openness, Conscientiousness, Extraversion, Agreeableness, and Neuroticism) as sufficient to discriminate between individual personalities.

In this work, we seek to explore two facets of users of multimedia social networks: (i) the correlations that exist between users' (demographic and psychological) profiles and the latent roles that emerge from their online behavior, and (ii) how the characteristics of broadcasts influence their popularity. We place this work in the context of related work (Section 2). We introduce our method of collecting Facebook user data, together with associated profiles (Section 3). In Section 4, we detail our use of clustering algorithms with user features to group similar individuals, and our characterization of the most successful broadcasts within our data set. We present (Section 5) and discuss (Section 6) our results from this analysis. Section 7 closes with our conclusions and our expectations for further work.

## 2. RELATED WORK

User behavior in online social networks has been well-studied, but the correlation of this behavior with formal roles within these networks has not yet (to our knowledge) been attempted. An early study [5] characterized user behavior in Usenet newsgroups, explicitly focusing on 'emergent' (latent) roles (such as question-poser and question-answerer) rather than structural (formal) roles.

Similarly, the authors of [6] presented models of the browsing habits of users across several social networks without explicit reference to information about the users themselves. A subset of YouTube users were segregated in [7] according to observed user features, each of these groupings defining a latent role within this community (but again without a corresponding formal role). In this work, we extend the approach from [7]: we group users according to features of their behavior on Facebook, but then also review the demographic and psychological data associated with each cluster to interpret the inferred formal roles. The authors of [2] defined roles of users in transactional data before the emergence of social media; our work aims to identify the analogous roles via a data-driven clustering approach to define behavior categories.

The social (and financial) rewards of generating 'viral' multimedia broadcasts have led to detailed investigation of the diffusion of information across social networks. Recent studies have explored the effect of content [8], and social and organizational context [9] on the popularity of text-based broadcasts. Various types of characteristics (such as video length) that top viral videos share were presented in [10]. We seek to identify and compare such viral characteristics across the *mixed* media types (text, photos, videos) found on Facebook.

## 3. DATA COLLECTION

**Table 1: Summary of Facebook data set**

| Type | Count |
|---|---|
| Users | 1327 |
| Videos | 4903 |
| Albums | 22658 |
| Photos | 289313 |
| Posts | 489929 |
| Post comments | 244809 |

In order to acquire a large Facebook dataset while still maintaining user privacy, we developed an extraction tool deployed as a web service to collect anonymized data from volunteers' Facebook accounts. We recruited 1327 participants from across the U.S. that were at least 18 years old. Although we employed emailing colleagues and posting on Facebook for recruiting, most (>90%) of our participants were not affiliated with our companies nor were they friends with any of the involved researchers. Upon signing up for the study and giving informed consent, participants first answered an online survey. In that survey, we collected participants' Big Five Personality scores using a questionnaire adapted from International Personality Item Pool [4], as well as their demographic information (such as age, gender, marital status and education). After that, participants authorized a Facebook application to collect their Facebook activity data (including the content of their Facebook 'Wall'). In order to ensure an efficient data collection, we placed (generous) caps on the amount of each data type kept. For example, we restricted data to be from the last 365 days, and accumulated at most 1000 posts from each user. Like other standard Facebook applications, this application uses OAuth, an open authorization standard to get permission from users, requiring neither passwords nor user names from participants. For privacy protection, study participants (and their Facebook friends) were assigned random unique IDs, and only these IDs are kept as an identifier in the extracted data set. In addition, we computed a set of high-level, aggregated statistics from each participant's activities. No raw text content or visual content from a user or the user's friends was collected. Instead, we replaced each distinct word occurring within text content with a unique integer. By these steps, we ensured that it is impossible to reconstruct personal identifiable information from users. Table 1 presents the quantities of different types of Facebook broadcasts (video, album, photo, post, comment) aggregated using our tool.

**Table 2: Summary of demographic profiles**

| Parameter | Percentage | Values |
|---|---|---|
| Age | | 18 – 71 years (Mean 29 years, S.D. 8 years) |
| Gender | 59% | Female |
| | 41% | Male |
| Marital Status | 52% | Single/divorced |
| | 48% | Engaged/married/partnered |
| Children | 64% | None |
| | 33% | Some that live at home |
| | 3% | None that live at home |
| Job Status | 40% | Full-time Worker |
| | 19% | Part-time Worker/Student |
| | 17% | Full-time Student |
| | 13% | Unemployed |
| | 10% | Home-maker |
| | 1% | Retired |
| Job Category | 11% | Computer/IT |
| | 8% | Education/Training |
| | 7% | Retail |
| | 6% | Medical/Health |
| | 68% | Other |
| Native English Speaker? | 96% | Yes |
| | 4% | No |

Not every participant provided a valid survey. We received complete surveys from only 1175 participants, and we tested the validity of these surveys using the time taken to complete the survey and the agreement between questions targeting the same personality trait. Table 2 lists various demographic statistics from the remaining 984 valid surveys, showing a diverse population of participants.

We expected few Facebook connections between our participants – we considered two users to be connected if they were listed as each other's friend or had posted on each other's wall. Instead we found our user group was well connected, and present this connectivity in Fig. 1, where we include only those connections and nodes required for the shortest paths between users.

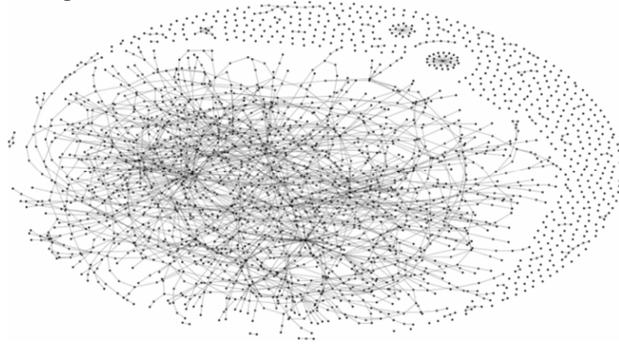

**Fig. 1: Network graph for our Facebook participants**

## 4. DATA ANALYSIS

We applied two separate analyses to our Facebook data set: (i) we clustered users based on behavior features to search for evidence of a correlation between user behavior and formal roles, and (ii) we measured the variation in broadcast popularity with properties of the broadcast in order to identify common characteristics of successful broadcasts.

### 4.1 Comparing profile data within user clusters

Our first goal is to find sets of behaviors that commonly co-occur and thus define categories of user types, thereby defining the latent roles users take on. Further, we desire the clusters to be data-driven, rather than attempting to match behaviors to a pre-defined set of roles, so we can discover the novel communication behaviors that emerge within social media. A common data-mining approach to extracting entity groupings (without any prior knowledge of the grouping characteristics) is the application of the standard K-means algorithm. Here we followed the example of the authors of [7] by identifying user features of interest, and computing the clustering of features. We then extended their approach by comparing the user profiles within each cluster.

We first applied a minimum activity threshold to users, and skipped users with fewer than 10 broadcasts over the time interval of our data set. Of the 1,327 original participants, 1,182 users passed this threshold. Note that not every active Facebook user had a corresponding survey, and not every survey had a corresponding active Facebook user, for reasons including this filtering and the above-mentioned survey validation. A further pre-processing step was necessary due to the ability of Facebook users to post to other users' walls: we excluded content authored by other users from the analysis of a given user.

As part of our analysis, we measured how the range of topics found in the text of post and multimedia broadcasts varied from one user to the next. Since we did not have access to the raw text, we measured this diversity using Latent Dirichlet Allocation (LDA) topics derived from the integer representation of words. We excluded from consideration the 100 most frequent words (based on a Wiktionary frequency list [11]), as well as short (less than 3 characters) or stop words. We also ignored those words not occurring more than 5 times in our overall corpus of 160,557 words. We were left with 34,565 words. We then applied LDA tools from LingPipe [12] separately on the anonymized text associated with posts and multimedia content, extracting 100 topics from each. Finally, for each broadcast, we calculated the three most probable of these LDA topics for the text associated with the broadcast. The diversity in topics across a user's post/multimedia broadcasts was then calculated as the number of distinct probable topics found across all of the user's post/multimedia broadcasts, respectively.

For the purposes of this study, we identified the following 12 user features as being of most interest:

i. **Average number of broadcasts per day ($f_1$)**: The number of wall posts, post comments, and video broadcasts authored by the user averaged over the number of days the user was active on Facebook. Photos do not have a broadcast time in our data set and so we excluded photo and album broadcasts from this count.
ii. **Number of friends ($f_2$)**: The total number of Facebook friends of the user.
iii. **Most typical time of day for broadcasts ($f_3$)**.
iv. **Ratio of private to public broadcasts ($f_4$)**: Facebook users can limit who can view a broadcast by selecting a privacy option. The default is to broadcast publicly. This feature compares the frequency of private broadcasts to that of public broadcasts.
v. **Relative frequency of *status*-type posts ($f_5$)**: The fraction of posts (that can be of type *status*, *check-in*, *photo broadcast*, *album broadcast*, *link broadcast*, *music broadcast*, *offer*, *question*, or *video broadcast*) that are of type *status*.
vi. **Diversity of topics within text posts ($f_6$)**.
vii. **Number of text broadcasts ($f_7$)**.
viii. **Number of video broadcasts ($f_8$)**.
ix. **Number of photo/album broadcasts ($f_9$)**.
x. **Average length of video ($f_{10}$)**.
xi. **Average length of video, photo and album captions ($f_{11}$)**.
xii. **Diversity of topics within video, photo and album captions and descriptions ($f_{12}$)**.

We computed the feature vector $[f_1, f_2, f_3, f_4, f_5, f_6, f_7, f_8, f_9, f_{10}, f_{11}, f_{12}]$ as input to our K-means algorithm. K-means clustering relies on equal scaling between vector elements. Since these features have different units, we first normalized the feature values so that each normalized feature ranged from 0 to 1. As part of this normalization, we identified and excluded data points with significantly outlying values that skewed the median feature value, and removed less than 0.5% of the original data set.

We then applied the K-means clustering algorithm to identify clusters of these feature vectors, and thus divide our Facebook users into groups of similar behavior. We employed the Weka [13] implementation of this algorithm, and used the standard Euclidean distance as the distance measure. For this initial analysis, we derived an optimal number of clusters heuristically; in future work, we will apply various techniques (such as the one described in [7]) to algorithmically compute this number, and also discuss the effect our selection of clustering algorithms has on our results. For users with null feature values (such as a user with no videos and thus a null average video length), our chosen clustering algorithm assigns the mean feature value. Instead of accepting this behavior, we instead assigned a default value (such as an average video length of 0).

For the users within each feature cluster, we then aggregated the corresponding demographic and psychological profiles, and identified common characteristics that were distinct to that cluster of users.

### 4.2 Effect of broadcast and user characteristics on propagation of Facebook broadcasts

The *success* of a Facebook broadcast can be measured by the number of *likes* and comments that that broadcast receives, as this number indicates the number of Facebook users engaged enough by the broadcast to interact with it. As part of our initial analysis of the successful broadcasts within our Facebook data set, we examined how success varied with each of the following broadcast characteristics:
  i. **The privacy settings of the broadcast**: Whether broadcast was private or public.
  ii. **The type of post**: Whether the post was a *status update*, *check-in*, etc. – see Section 4.1.
  iii. **The type of broadcast**: Was it a post, comment, album broadcast or photo broadcast.
  iv. **The presence of user *tags* within photos**: Whether a photo broadcast contained *tags* (labels) for any users present in that photo.

## 5. RESULTS

### 5.1 Latent roles within Facebook

The five user clusters derived from our Facebook data are shown in Fig. 2, with histograms representing the (normalized) feature values for each cluster centroid. The line diagram indicates the numbers of users within each cluster – thus we see that Cluster 4 is the largest.

We note that the most distinguishing features for the clusters are the relative frequency of status-type posts ($f_5$), the diversity of topics within text posts ($f_6$), and the topic diversity within video and photo captions ($f_{12}$). We also point out that the mean value of several features ($f_2, f_3, f_{11}$) do not vary much across clusters. The overall distribution of these three features shows there is a wide range in values, but most users have a similar value for each of the three features. Clearly, the users with outlying values did not share enough other distinct characteristics for the K-means algorithm to require additional clusters. For each feature, we applied to each pair of cluster distributions the Kolmogorov-Smirnov test (for continuous-valued features) or Pearson's Chi-squared test (for discrete-valued features) to confirm the significance of the cluster distributions.

Given these five clusters, we examined how the 12 features, as well as the demographic and psychological parameters, varied for each cluster. For example, Fig. 3 shows the different distributions in perceived stress, from which we can deduce that the members of Cluster 2 have

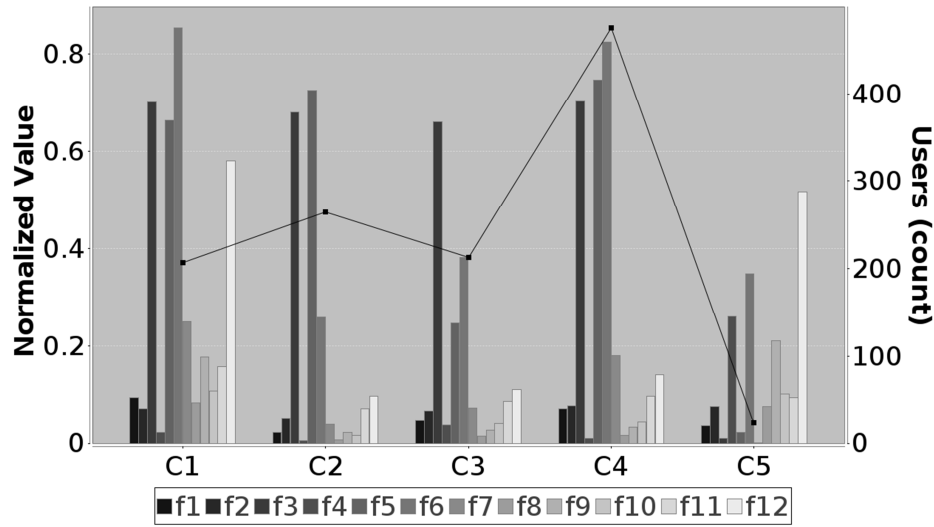

**Fig. 2: (histogram) Users clustered by 12 Facebook activity features using K-means algorithm; (line plot) Number of users within clusters**

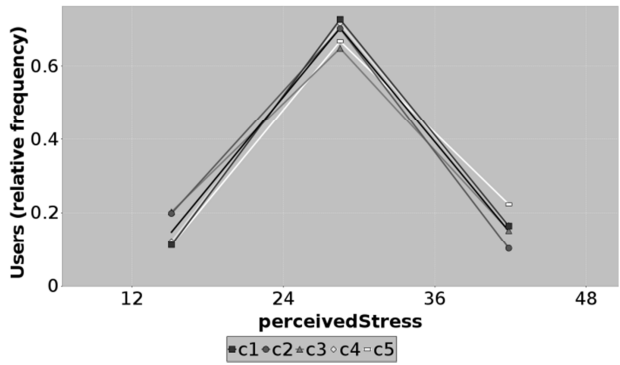

**Fig. 3: Relative frequency of users that have low, medium and high levels of perceived stress for our overall population (*solid black line*) and for the five clusters (*symbols*)**

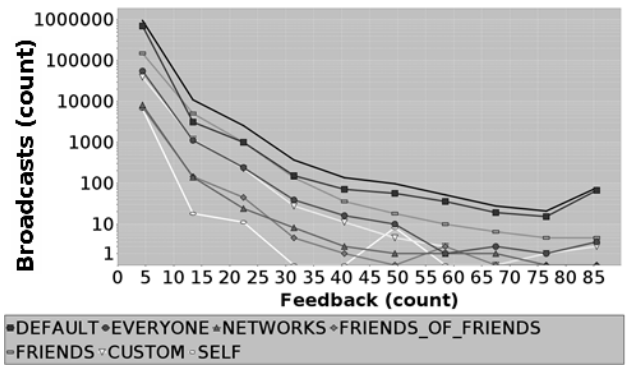

**Fig. 4: Number of posts with a given privacy setting that received likes or comments as a function of the amount of such feedback**

lower stress levels, in general. We summarize the derived characteristics of the five clusters as follows:

i. **Cluster 1 – Multimedia-Savvy & Engaged Users:** This cluster represents the multimedia super-broadcasters, with large numbers of broadcasts of every type, and a consistently high topic diversity for posts and multimedia broadcasts. The users are mostly female. Contrary to our expectations (due to their engagement), they have a normal range in extraversion scores. They have higher neuroticism and conscientiousness scores.

ii. **Cluster 2 – Low Engagement Users:** These users are not very involved in the Facebook network. They have a low number of broadcasts of every type, and tend to broadcast publicly. This group also has the lowest topic diversity in posts. They have higher levels of education and self-assurance, and lower stress levels.

iii. **Cluster 3 – Private broadcasters:** This small group of users has a diverse range in broadcast types and broadcast a smaller number of broadcasts less publicly. The cluster is predominantly older and employed full-time, with lower stress, extraversion and excitement seeking scores, and higher overall mood scores.

iv. **Cluster 4 – High Engagement Users:** This, the largest cluster, represents the super-broadcasters. This cluster is similar to Cluster 1, except that this group is mostly interested in text broadcasts, and has low multimedia broadcast numbers and topic diversity. The users are mostly young, and share more publicly than Cluster 1. Again contrary to our expectations, they have a normal range in extraversion scores.

v. **Cluster 5 – Multimedia Specialists:** This is a small cluster with users who privately broadcast only multimedia with correspondingly high topic diversity. The group is mostly young and female, mostly stay-at-home or unemployed, and with a less-than-average education level.

## 5.2 Characteristics of popular broadcasts

Fig. 4 shows the binned feedback (likes and comments) counts associated with posts of different privacy settings in our data sets. The number and size of feedback bins were selected to cover all the data except for some outliers. These outliers had much higher feedback values ($\cong 10,000$) and the *default* privacy setting, and have been included in the highest value feedback bin. From this figure we see that, while high feedback values are associated with every privacy setting, the most popular broadcasts are mostly *default* or *friends-only* broadcasts.

Fig. 5 shows the feedback distributions for the different types of posts, with a similar outlier exclusion. Posts sharing a photo are the most popular, followed by status updates.

We examined which types of broadcast (post, comment, album/photo broadcast) received the most feedback. All but comments received the highest levels of feedback; and album broadcasts were the most popular. Although we expect *tagging* users within photos increases feedback, the most popular photo broadcasts contained no tags – and likely were not photos of users – which might indicate the subjects of these photos had a more general appeal than photos of specific users.

## 6. DISCUSSION

This work shows the promise of our approach to user modeling. Using straightforward features easily derived from Facebook user data, we have identified latent roles, or types of behavior patterns that individuals follow. Further, we have begun to analyze how those latent roles correlate with other observable participant features, such as formal roles and demographics data. We have seen how expectations derived from offline experience (such as high extraversion scores for gregarious users) are not necessarily met online. One of the benefits of such user role analysis in social media is that it enables us to interpret people's behaviors in internal social media forums, as well as open forums. This information could support insider threat detection because it can help: (1) flag users that behave inconsistently compared to their "peers" according to demographics or formal roles; and (2) provide another measure of expected individual behavior, from which we can identify when individuals adopt new behavior patterns.

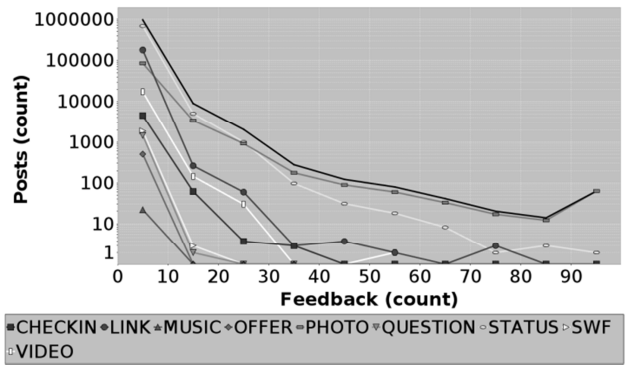

**Fig. 5: Number of posts of a given type as a function of the feedback received**

From our analysis, it is clear that there are two types of multimedia broadcasts that generate the most feedback: (i) public broadcasts that expose the content to the greatest audience and thus maximize the possibility of reaching an interested party, and (ii) broadcasts of private content with close friends that maximize engagement by including content of most relevance to these friends.

In future work, we will perform a complimentary analysis of a Gmail dataset and a comparison with this Facebook dataset, to see if the results from this paper hold across online social communication platforms, and to therefore draw deeper insights into online behaviors. Having derived latent roles within this social network, we will seek to determine the impact of these roles on networks. In particular, we will combine our two research directions by examining how success varies with cluster membership, and whether the multimedia properties or the user characteristics have greater influence on the popularity of a broadcast.

## 7. CONCLUSIONS

We presented a unique extension of prior work on clustering social network users that attempts to correlate the latent roles associated with these clusters with demographic and psychological profiles. We also listed properties common to the most popular broadcasts. Future work will seek to identify anomalous behavior on limited-access networks, and to examine whether the author or the content of a broadcast drives its popularity.

## ACKNOWLEDGEMENTS

The authors gratefully acknowledge support for this work from DARPA through the SMIITE (Social Multimedia Investigation for Insider Threat Exposure) and GLAD-PC (Graph Learning for Anomaly Detection using Psychological Context) projects funded as part of the ADAMS (Anomaly Detection At Multiple Scales) program. Any opinions, findings, and conclusions or recommendations in this material are those of the authors and do not necessarily reflect the views of the government funding agencies.